\documentclass[journal,twoside, print]{ieeecolor}
\usepackage{generic}
\usepackage{cite}
\usepackage{amsmath,amssymb,amsfonts}
\usepackage{algorithmic}
\usepackage{graphicx}
\usepackage{algorithm,algorithmic}
\usepackage{hyperref}
\hypersetup{hidelinks=true}
\usepackage{textcomp}
\usepackage{array,multirow}
\usepackage{adjustbox}

\newcommand{\add}[1] {\textcolor{black}{#1}} 
\newcolumntype{P}[1]{>{\centering\arraybackslash}p{#1}}
\newcolumntype{M}[1]{>{\centering\arraybackslash}m{#1}}

\def\BibTeX{{\rm B\kern-.05em{\sc i\kern-.025em b}\kern-.08em
    T\kern-.1667em\lower.7ex\hbox{E}\kern-.125emX}}
\begin{document}
\title{Classification of Multi-Parametric Body MRI Series Using Deep Learning}
\author{Boah Kim, Tejas Sudharshan Mathai, Kimberly Helm, Peter A. Pinto, Ronald M. Summers
\thanks{This research was supported in part by the Intramural Research Program of the National Institutes of Health, Clinical Center, United States, and in part by the Center for Cancer Research, National Cancer Institute, United States. }
\thanks{This work expands on our previous study \cite{kim2024automated} presented at the 21st IEEE International Symposium on Biomedical Imaging (ISBI 2024).}
\thanks{Boah Kim, Tejas Sudharshan Mathai, Kimberly Helm, and Ronald M. Summers are with Radiology and Imaging Sciences, National Institutes of Health Clinical Center,  
Bethesda, MD 20892 USA (e-mail: boah.kim@nih.gov; tejas.mathai@nih.gov; kimhelm@seas.upenn.edu; rms@nih.gov). }
\thanks{Peter A. Pinto is with National Cancer Institute, Bethesda, MD 20892 USA (e-mail: pintop@mail.nih.gov).}
}

\maketitle

\begin{abstract}
Multi-parametric magnetic resonance imaging (mpMRI) exams have various series types acquired with different imaging protocols. The DICOM headers of these series often have incorrect information due to the sheer diversity of protocols and occasional technologist errors. To address this, we present a deep learning-based classification model to classify 8 different body mpMRI series types so that radiologists read the exams efficiently. Using mpMRI data from various institutions, multiple deep learning-based classifiers of ResNet, EfficientNet, and DenseNet are trained to classify 8 different MRI series, and their performance is compared. Then, the best-performing classifier is identified, and its classification capability under the setting of different training data quantities is studied. Also, the model is evaluated on the out-of-training-distribution datasets. Moreover, the model is trained using mpMRI exams obtained from different scanners in two training strategies, and its performance is tested. Experimental results show that the DenseNet-121 model achieves the highest F1-score and accuracy of 0.966 and 0.972 over the other classification models with p-value$<$0.05. The model shows greater than 0.95 accuracy when trained with over 729 studies of the training data, whose performance improves as the training data quantities grew larger. On the external data with the DLDS and CPTAC-UCEC datasets, the model yields 0.872 and 0.810 accuracy for each. These results indicate that in both the internal and external datasets, the DenseNet-121 model attains high accuracy for the task of classifying 8 body MRI series types. 
\end{abstract}


\section{Introduction}
\label{sec:introduction}

\begin{figure}[!t]
\centering
\centerline{\includegraphics[width=\linewidth]{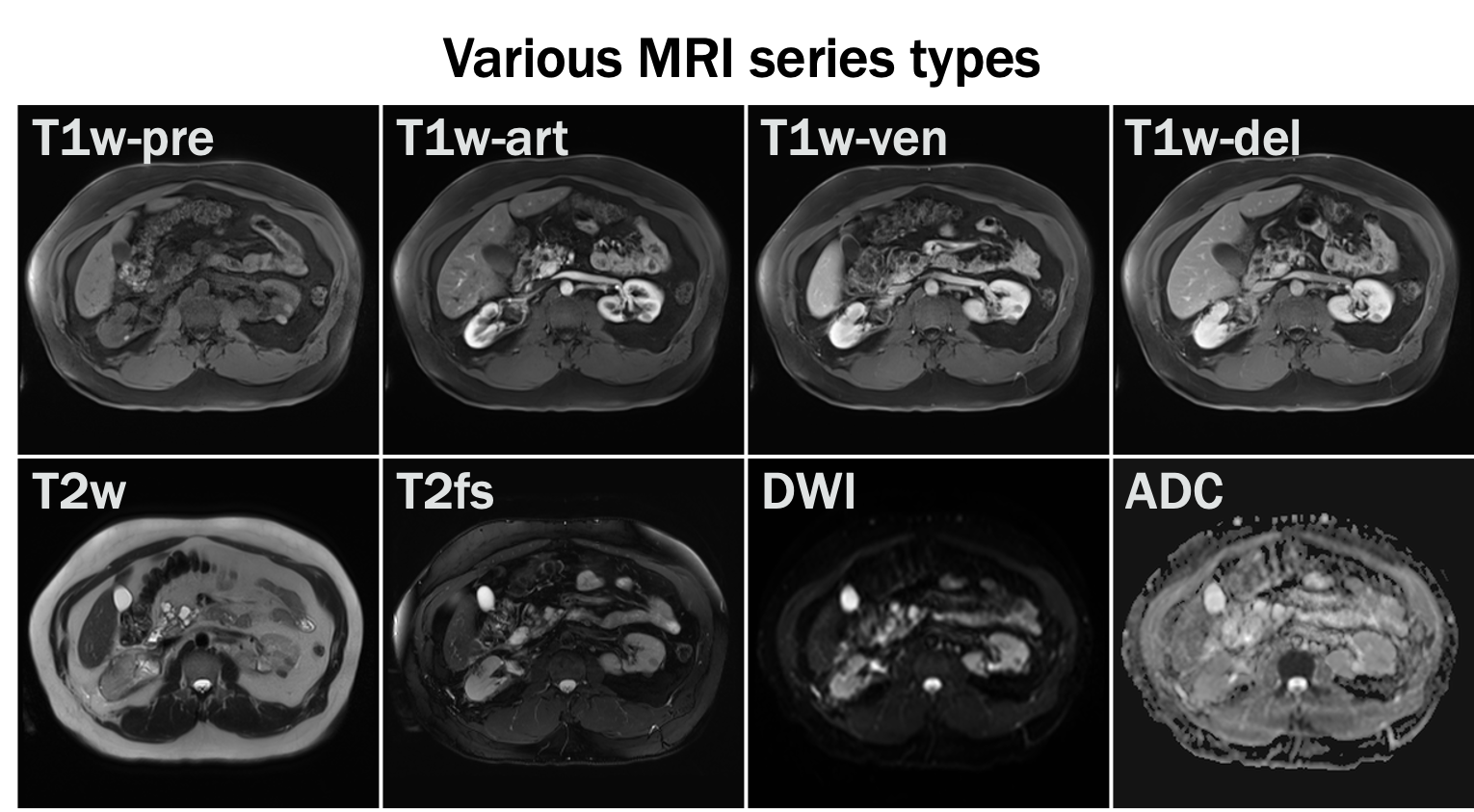}}
\caption{Examples of the 8 different types of body MRI series: pre-contrast T1-weighted (T1w-pre), contrast-enhanced T1-weighted in the arterial (T1w-art), portal venous (T1w-ven), and delayed (T1w-del) phases, T2-weighted (T2w), fat-suppressed T2-weighted (T2fs), diffusion-weighted imaging (DWI), and derived apparent diffusion coefficient (ADC) series. }
\label{fig: intro}
\end{figure}

\IEEEPARstart{M}{ulti}-parametric magnetic resonance imaging (mpMRI), acquired using different echo times, repetition times, radio-frequency pulses, and other parameters, shows different image features and pathological information as shown in Fig.~\ref{fig: intro}. This is a widely used imaging technique to diagnose various diseases. In particular, certain series combinations are frequently used since they provide complementary information related to disease status. For example, T2-weighted and diffusion-weighted imaging (DWI) allow the prostate zones to be diagnosed. Typically, the series description and other pertinent information related to the sequence are stored in the DICOM header, and logic-based rules are applied to the DICOM fields to arrange the mpMRI series in a picture archiving and communication system (PACS) viewer. This constitutes the hanging protocol for radiologists.  

However, a variety of MRI manufacturers, diverse protocols across different institutions, and subjective preferences of the MRI technologist at the time of image acquisition lead to heterogeneous, inconsistent, and subjective information in about 16\% of the DICOM header fields \cite{3_gueld2002quality, 4_anand2023automated, 5_zhu20203d, 6_baumgartner2023metadata}. Such inaccuracies impede the use of the DICOM tags for automatic series categorization \cite{3_gueld2002quality}. Accordingly, radiologists often rearrange the series to read the exam. To reduce their oversight, an automated approach to classify the MRI sequence types could improve reading efficiency.

A majority of prior deep learning approaches have mostly focused on classifying different series in brain MRI studies \cite{7_liang2021magnetic, 8_ranjbar2020deep, 9_de2021deep, 10_noguchi2018artificial, 11_chou2022automatic}. There are limited works targeting body MRI classification and they have focused on a specific anatomy, such as the liver \cite{5_zhu20203d} or the prostate \cite{6_baumgartner2023metadata}. Zhu et al. \cite{5_zhu20203d} classified various liver MRI series with an accuracy of 0.82, while Baumgartner et al. \cite{6_baumgartner2023metadata} classified prostate MRI sequences with an accuracy of 0.99. In a prior study \cite{25_helm2024automated}, we were the first to propose an approach to classify 5 different MRI series obtained at the level of the chest, body, and pelvis. To the best of our knowledge, there is no other work to classify MRI series across the body regions.

In this work, we developed a framework to automatically classify 8 distinct series in mpMRI studies acquired at the level of the chest, abdomen, and pelvis. Here, our approach distinguished between T1-weighted (pre, arterial, portal venous, delayed phases), T2-weighted, fat-suppressed T2, DWI, and apparent diffusion coefficient maps. We first trained various learning-based classifiers on studies acquired by Siemens scanners. Then, the classifier with the highest performance was identified and trained further on studies from both Siemens and Philips scanners. The evaluation was conducted on held-out training data including different scanner datasets and two external test datasets.

\begin{figure*}[!t]
\centering
\centerline{\includegraphics[width=0.85\linewidth]{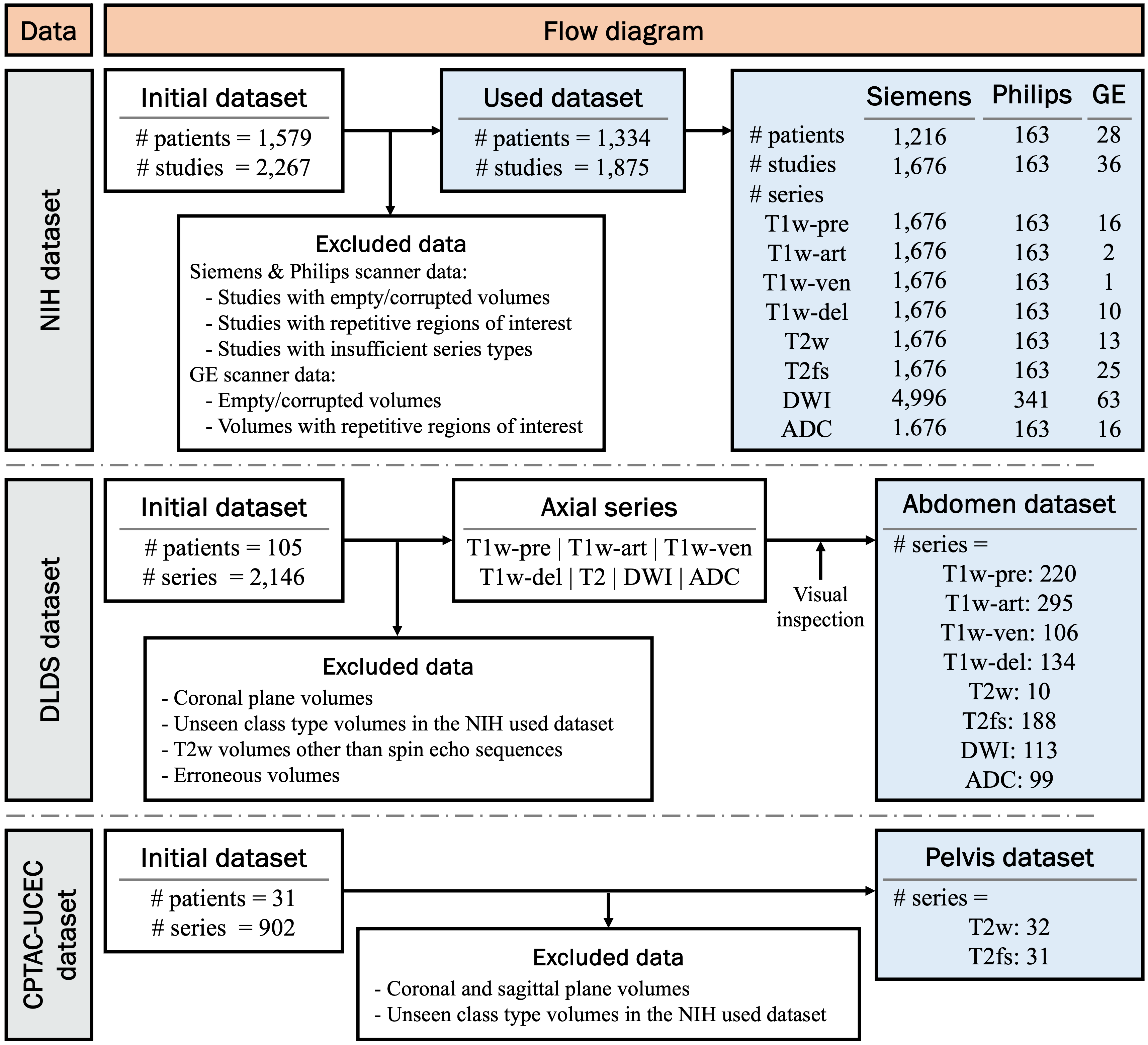}}
\caption{Standards for Reporting Diagnostic Accuracy (STARD) chart of the data collection process for the National Institutes of Health dataset (NIH) (top), the Duke Liver Dataset (DLDS) (middle), and the Clinical Proteomic Tumor Analysis Consortium Uterine Corpus Endometrial Carcinoma (CPTAC-UCEC) dataset (bottom).}
\label{fig: dataflow}
\end{figure*}

\begin{table*}[!t]
\centering
\caption{Demographics of patient studies for each dataset used for body MRI classification}
\label{tab:data}
\begin{tabular}{l|M{0.75cm}M{0.8cm}M{1.2cm}M{1.1cm}M{1.1cm}M{1.1cm}ccM{1.0cm}M{0.9cm}}
\hline
\multicolumn{11}{l}{\textbf{NIH dataset}} \\
\hline
\textbf{Dataset}         & \multicolumn{2}{c}{\textbf{\# patients}} & \multicolumn{2}{c}{\textbf{\# studies}}            & \multicolumn{2}{c}{\textbf{\# series}} & \multicolumn{2}{c}{\textbf{Gender ratio (Male:Female)}} & \multicolumn{2}{c}{\textbf{Age range (years)}}  \\ \hline
Siemens dataset & \multicolumn{2}{c}{1,216}       & \multicolumn{2}{c}{1,676}                 & \multicolumn{2}{c}{16,728}    & \multicolumn{2}{c}{614:602}                    & \multicolumn{2}{c}{4–85 (median=51)} \\
Philips dataset & \multicolumn{2}{c}{163}         & \multicolumn{2}{c}{163}                   & \multicolumn{2}{c}{1,482}     & \multicolumn{2}{c}{72:91}                      & \multicolumn{2}{c}{13-83 (median=50)}  \\
GE dataset & \multicolumn{2}{c}{28}       & \multicolumn{2}{c}{36}                 & \multicolumn{2}{c}{146}    & \multicolumn{2}{c}{12:16}                    & \multicolumn{2}{c}{12-80 (median=43)}  \\
\hline
\textbf{Series}          & \textbf{No. of scans}   & \textbf{No. of slices}  & \textbf{Slice thickness} (mm) & \textbf{Pixel spacing} (mm) & \textbf{No. of rows}  & \textbf{No. of columns} & \textbf{TR} (msec) & \textbf{TE} (msec) & \textbf{Flip angle} ($\circ$)   & \textbf{Field strength} (T)  \\ \hline
T1w-pre            & 1,855          & 54-144         & 2.5-6.0              & 0.37-1.56          & 256-1024      & 192-1024        & 3.11-7.39               & 1.43-3.50            & 9-15             & 1.5, 3.0            \\
T1w-art             & 1,843          & 60-140         & 2.5-6.0              & 0.53-1.56          & 256-640      & 192-560        & 3.11-6.81               & 1.43-3.13            & 9-12             & 1.5, 3.0            \\
T1w-ven             & 1,840          & 60-120         & 2.5-6.0              & 0.53-1.56          & 256-640      & 192-560        & 3.11-6.81               & 1.43-3.13           & 9-12             & 1.5, 3.0            \\
T1w-del             & 1,849          & 60-144         & 2.5-6.0              & 0.53-1.56          & 256-640      & 192-560        & 3.11-7.23               & 1.43-3.28            & 9-15             & 1.5, 3.0            \\
T2w         & 1,852          & 16-304          & 1.2-8.0              & 0.52-1.95          & 256-768      & 176-768        & 567.52-6190.28         & 78.85-241.47           & 90-180           & 1.5, 3.0            \\
T2fs            & 1,864          & 20-90          & 3.0-8.0              & 0.47-1.95          & 256-768      & 192-672        & 1263.6-18177.3          & 60.00-194.00           & 90-160           & 1.5, 3.0            \\
DWI             & 5,400          & 15-64          & 4.0-10.0              & 0.88-3.13          & 128-448      & 96-448         & 2400.0-17142.9           & 42.60-100.20          & 90               & 1.5, 3.0            \\
ADC             & 1,856          & 15-71          & 4.0-8.0              & 0.88-3.13          & 128-448      & 96-448         & 2000.0-16000.0           & 48.62-100.20          & 90               & 1.5, 3.0            \\
\hline \hline
\multicolumn{11}{l}{\textbf{DLDS dataset}}   \\
\hline
\textbf{Dataset}         & \multicolumn{2}{c}{\textbf{\# patients}} & \multicolumn{2}{c}{\textbf{\# studies}}            & \multicolumn{2}{c}{\textbf{\# series}} & \multicolumn{2}{c}{\textbf{Gender ratio (Male:Female)}} & \multicolumn{2}{c}{\textbf{Age range (years)}}  \\ \hline
Abdomen dataset & \multicolumn{2}{c}{105}         & \multicolumn{2}{c}{-}                     & \multicolumn{2}{c}{1,165}     & \multicolumn{2}{c}{76:29}                      & \multicolumn{2}{c}{30-83 (median=62)}  \\ \hline
\textbf{Series}          & \textbf{No. of scans}   & \textbf{No. of slices}  & \textbf{Slice thickness} (mm) & \textbf{Pixel spacing} (mm) & \textbf{No. of rows}  & \textbf{No. of columns} & \textbf{TR} (msec) & \textbf{TE} (msec) & \textbf{Flip angle} ($\circ$)   & \textbf{Field strength} (T)  \\ \hline
T1w-pre         & 220            & 48-100         & 3.0-7.0              & 0.68-1.76          & 168-576      & 256-576        & 3.46-9.20               & 1.07-3.126           & 4-15         & 1.5, 3.0            \\
T1w-art         & 295            & 48-168         & 3.0-7.0              & 0.68-1.56          & 208-512      & 256-512        & 2.83-6.96               & 1.28-3.268           & 9-12         & 1.5, 3.0            \\
T1w-ven         & 106            & 44-100         & 3.0-7.0              & 0.68-1.56          & 208-512      & 256-512        & 4.00-6.78               & 1.23-3.268           & 9-12         & 1.5, 3.0            \\
T1w-del         & 134            & 30-205         & 3.0-7.0              & 0.70-1.56          & 208-512      & 256-512        & 3.66-124.00             & 1.07-3.126           & 9-70         & 1.5, 3.0            \\
T2w             & 10             & 15-42          & 5.0-8.0              & 0.63-1.56          & 250-640      & 256-640        & 561.02-6891.25          & 78.0-98.784          & 90-167       & 1.5, 3.0            \\
T2fs            & 188            & 20-46          & 5.0-8.0              & 0.63-1.64          & 208-640      & 256-640        & 700.0-17142.9           & 75.936-117.0         & 90-180       & 1.5, 3.0            \\
DWI             & 113            & 20-49          & 6.0-8.0              & 0.94-2.40          & 126-384      & 192-384        & 3000.0-8200.0           & 55.2-82.0            & 90             & 1.5, 3.0            \\
ADC             & 99             & 27-49          & 6.0                  & 0.94-2.40          & 126-384      & 192-384        & 4000.0-8200.0           & 55.2-82.0            & 90             & 1.5, 3.0            \\
\hline \hline
\multicolumn{11}{l}{\textbf{CPTAC-UCEC dataset}} \\
\hline
\textbf{Dataset}         & \multicolumn{2}{c}{\textbf{\# patients}} & \multicolumn{2}{c}{\textbf{\# studies}}            & \multicolumn{2}{c}{\textbf{\# series}} & \multicolumn{2}{c}{\textbf{Gender ratio (Male:Female)}} & \multicolumn{2}{c}{\textbf{Age range (years)}}  \\ \hline
Pelvis dataset  & \multicolumn{2}{c}{31}          & \multicolumn{2}{c}{31}                    & \multicolumn{2}{c}{63}        & \multicolumn{2}{c}{0:31}                       & \multicolumn{2}{c}{41-73 (median=67)}  \\ \hline
\textbf{Series}          & \textbf{No. of scans}   & \textbf{No. of slices}  & \textbf{Slice thickness} (mm) & \textbf{Pixel spacing} (mm) & \textbf{No. of rows}  & \textbf{No. of columns} & \textbf{TR} (msec) & \textbf{TE} (msec) & \textbf{Flip angle} ($\circ$)   & \textbf{Field strength} (T)  \\ \hline
T2w             & 32             & 28-43          & 4.0                  & 0.78-1.09          & 320          & 320            & 2880.0-9820.0           & 119.0-128.0          & 125-127      & 1.5                 \\
T2fs            & 31             & 28-43          & 4.0                  & 0.78-1.09          & 320          & 320            & 3190.0-9250.0           & 98.0-105.0           & 127            & 1.5       \\
\hline
\multicolumn{11}{l}{
\textit{Note.} T1w = T1-weighted imaging with all phases, TR = repetition time, TE = echo time, “-“ indicates an unknown number. }
\end{tabular}
\end{table*}

\begin{figure*}[!t]
\centering
\centerline{\includegraphics[width=0.9\linewidth]{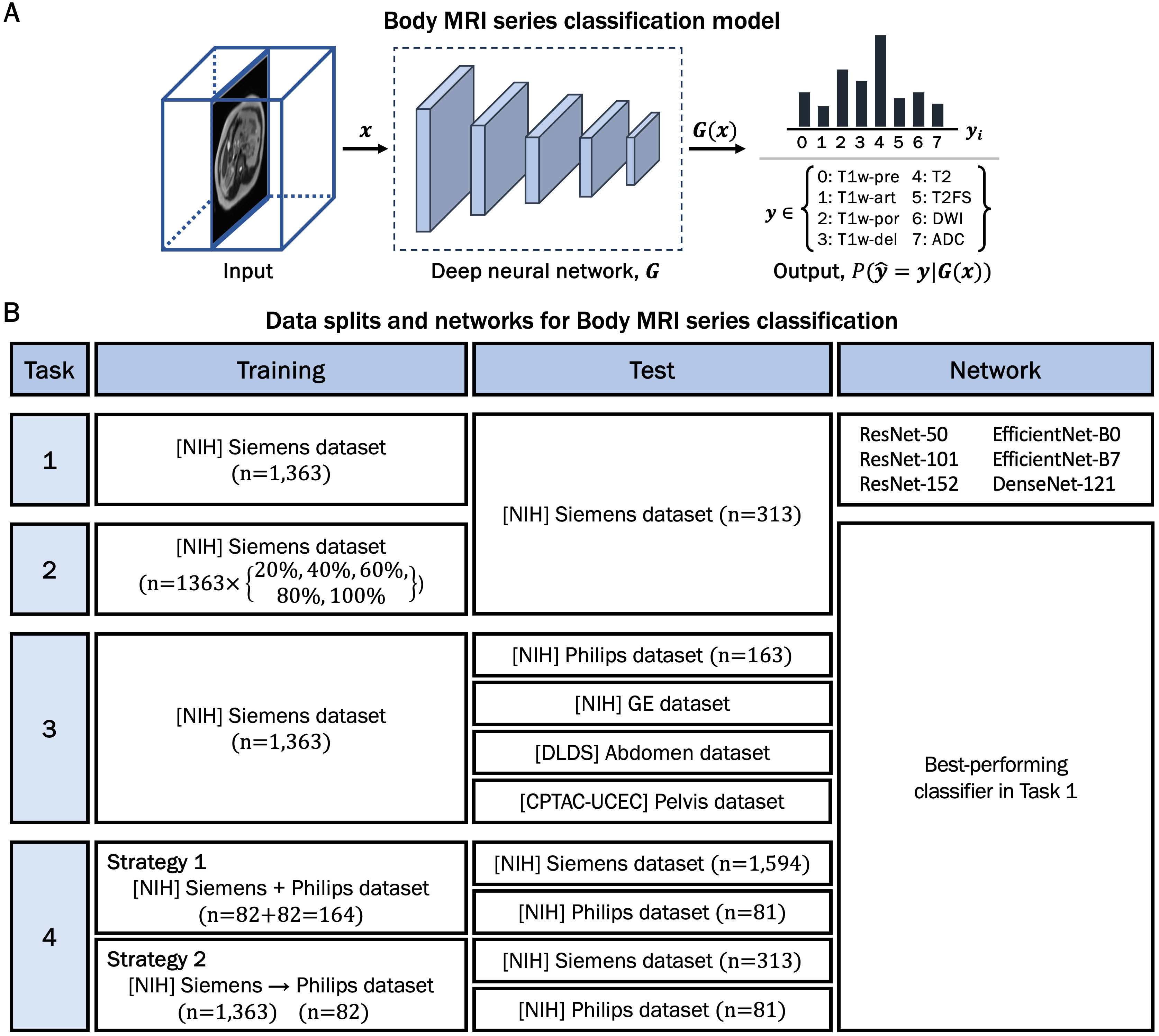}}
\caption{Overview of our work. (A) The framework for body MRI series classification model. A deep neural network takes 3D volumes as input and outputs probability scores for each series type. (B) Data splits and networks for 4 tasks in our work. Task 1 is to compare various classification networks, Task 2 is to study the best-performing classifier by varying the amount of training data, Task 3 is to evaluate the model on the external out-of-training distribution data, and Task 4 is to study the model training with data from different scanners. [NIH] is the National Institutes of Health dataset, [DLDS] is the Duke Liver Dataset, and [CPTAC-UCEC] is the Clinical Proteomic Tumor Analysis Consortium Uterine Corpus Endometrial Carcinoma dataset. The number of studies for each training and test data of the NIH Siemens and Philips dataset is indicated in parentheses. }
\label{fig: pipeline}
\end{figure*}

\section{Backgrounds}

\subsection{\add{Multi-parametric MRI}}
\add{Multi-parametric MRI (mpMRI) is one of the main imaging modalities to detect and characterize lesions. This combines morphological and functional information \cite{giannini2015fully} according to different parameters such as repetition time (TR) and time to echo (TE). TR is the time between successive pulse sequences and TE is the time between delivery of the radio frequency pulse and reception of the echo signal. In the following, among various MRI series types, we briefly introduced the commonly used series - T1-weighted, T2-weighted, fat-suppressed T2, diffusion-weighted imaging (DWI), and apparent diffusion coefficient map (ADC) - that we utilized in this work. 
}

\add{First, the T1-weighted MRI enhances the differences in longitudinal relaxation times (T1) of tissues, where T1 is the time taken for excited protons to return to equilibrium \cite{bushberg2011essential}. Using short TR and TE times, the T1-weighted images emphasize the fatty tissue signal and suppress the water signal. Thus, fat tissues appear bright whereas fluid-filled structures such as the bladder and ureter appear dark. Here, if the contrast agent is injected, the images can be categorized into pre, arterial, portal venous, and delayed phase series according to the injection time.}

\add{Second, the T2-weighted MRI strengthens the differences in the transverse relaxation times (T2) of tissues, in which T2 is the time taken for excited protons to lose phase coherence with each other \cite{bushberg2011essential}. The T2-weighted images are obtained using long TR and TE times and enhance the water signal. Therefore, the anatomical structures filled with fluids appear bright.}

\add{Third, the fat-suppressed T2 MRI is the T2-weighted image applying fat suppression techniques \cite{delfaut1999fat}. The fat suppression makes the bright signal of fatty tissues dark so that the contrast between pathological changes related to fluid area and the adjacent fatty tissues is enhanced. This MRI series is useful for the visualization of lesions by differentiating them from soft tissues.}

\add{Fourth, diffusion-weighted imaging (DWI) is an MRI technique using characteristics of the diffusion of water molecules \cite{saito2016diffusion}. The DWI acquires images based on the b value which indicates how much the signal will be attenuated by diffusion. In tissues with high water components that have active diffusion motion, such as cysts and necrotic tissues, the image signal intensity decreases significantly as the b value increases, whereas in tissues with suppressed diffusion motion, such as solid tumors, the signal becomes less weak even if the b value is increased, leading to enhanced contrast between the lesions and the surrounding tissues.}

\add{Lastly, the apparent diffusion coefficient (ADC) image, which is a measure of the magnitude of diffusion within tissues, is obtained using multiple DWI images with various b values \cite{le2013apparent}. The change of the signal is proportional to the rate of diffusion, and the anatomical structures with a smaller magnitude of diffusion in the ADC images appear darker.}

\subsection{\add{Image Classification with Deep Learning}} 
\add{Deep learning approaches have been widely applied to various computer vision tasks of not only natural images \cite{ruby2020binary, bateni2020improved, gulzar2023fruit} but also medical images \cite{cai2020review, krishnapriya2023pre, jiang2023review}. In particular, convolutional neural networks (CNN) \cite{yamashita2018convolutional} have been used as a workhorse of image processing due to their excellent capability to extract feature extractions of images. For the image classification, there are several representative models \cite{krizhevsky2012imagenet, simonyan2014very, 17_he2016deep, 19_huang2017densely, 18_tan2019efficientnet} using CNN.}

\add{Specifically, traditional neural networks using CNN for image classification \cite{krizhevsky2012imagenet, simonyan2014very} are composed of convolutional blocks and a fully connected layer. They feed the previous layer's output directly to the next layer. Through the convolutional layers and pooling layers in each convolutional block, the network extracts the key image features by reducing the feature map size. Then, the final fully connected layer outputs the probabilities for each class to classify input images.}

\add{Upon the traditional network architectures, ResNet \cite{17_he2016deep} is built by introducing the residual block which the feature map of the previous layer is concatenated to the next layer's feature map. This can solve the gradient vanishing problem of the traditional CNN models that occurs when the number of layers increases. Also, DenseNet \cite{19_huang2017densely} enhances ResNet by concatenating all feature maps of previous layers to the next layer, which allows the network to get the initial feature map information in each layer and achieve better classification performance. In addition, EfficientNet \cite{18_tan2019efficientnet} uses a compound scaling that scales three dimensions of the network - width, depth, and resolution. By balancing these three dimensions, the model improves the performance while reducing the number of learnable parameters.}

\add{Although these networks were designed using natural images, they have shown successive performance in medical image classification such as automated disease diagnosis \cite{sarwinda2021deep, chauhan2021optimization, raza2023lung}. In this work, we employed 3D versions of the classification models to classify body MRI series acquired at the level of chest, abdomen, and pelvis. Particularly, while the high performance of 2D medical image classification is often achieved by adopting the models pre-trained on natural images, in our work, we showed high and stable classification performance by training the 3D models from scratch using thousands of MRI volumes.}

\section{Method}

\subsection{Study Patients and Data}
In this work, an mpMRI dataset collected at the National Institutes of Health (NIH) was used, which was Health Insurance Portability and Accountability Act compliant and approved by the Institutional Review Board at the NIH. The publicly available Duke Liver Dataset (DLDS) \cite{12_macdonald2023duke}, and the Clinical Proteomic Tumor Analysis Consortium Uterine Corpus Endometrial Carcinoma (CPTAC-UCEC) dataset \cite{13_cptac} were also used. In this paper, the axial sequences were of primary interest. For all datasets, the requirement for signed informed consent from the patients was waived. Fig. \ref{fig: dataflow} shows the data collection process for each dataset, and Table~\ref{tab:data} provides detailed scan parameters of each series. 

\subsubsection{National Institutes of Health dataset (NIH)}
In the NIH dataset, 1,579 patients who had undergone MRI exams of their chest, abdomen, and pelvis between January 2015 and September 2019 were identified \cite{14_mathai2024universal, 15_mathai2023universal}. The MRI studies that imaged regions outside the body such as the head, neck, and extremities were excluded. Studies were also excluded if any series in the study was empty, contained repeated slices, or was corrupted in some manner. Through this process, 1,875 studies from 1,334 patients were gathered. We collected the following 8 axially imaged series: unenhanced T1-weighted (T1w-pre), contrast-enhanced T1-weighted in the arterial (T1w-art), portal venous (T1w-ven), and delayed (T1w-del) phases, T2-weighted (T2w), fat-suppressed T2 (T2fs), diffusion-weighted imaging (DWI), and derived apparent diffusion coefficient (ADC) maps. Here, we used all the available DWI series obtained with low (0–200 $s/mm^2$), intermediate (400–800 $s/mm^2$), and high (800–1,400 $s/mm^2$) b-values, where the DWI data with different b-values in each study had the same voxel spacing. 


In this dataset, 1,676 out of 1,839 studies were acquired with three Siemens scanners (Aera, Verio, and BioGraph mMR), 163 studies were obtained on a Philips scanner (Achieva), and the remaining 28 studies were scanned on three GE scanners (Signa, Optima, and Discovery). Each study in the Siemens and Philips scanner datasets contains all 8 types of series data, where all series data were resampled to match the volumetric dimensions of the DWI series \cite{brett_2023_7795644} as the DWI series had the smallest spatial dimensions. The studies from the Siemens scanner covered the chest (n=32), chest and abdomen (n=1), abdomen (n=1,573), abdomen and pelvis (n=18), and pelvis (n=52), and all the studies from the Philips scanner were obtained at the abdomen level (n=163).
On the other hand, each study in the GE scanner dataset contains at least one of the 8 types of series data, and the studies covered the abdomen (n=32), and abdomen and pelvis (n=4). The Siemens dataset was used for training and testing by splitting the data into 1,363 and 313 studies, respectively, and the Philips and GE datasets were used only for testing.

\subsubsection{Duke Liver Dataset (DLDS)}
The external DLDS dataset \cite{12_macdonald2023duke} provided 2,146 abdominal MRI series from 105 patients acquired at Duke University with 17 different series types. The MRI exams in this dataset were acquired to identify hepatic cirrhosis. The imaging protocols have been described in previously published literature \cite{12_macdonald2023duke}. In this work, based on series description, axial T1w-pre, T1w-art, T1w-ven, T1w-del, T2w, DWI, and ADC series were used. We combined the early-, mid-, and late-arterial T1w series into one T1w-art class as these were adjudged by a board-certified radiologist (R.M.S.) with 30+ years of experience to have been acquired with very little time difference between each other \cite{5_zhu20203d, 12_macdonald2023duke}. Also, since the DLDS dataset included the T2fs series in the axial T2w series, the same radiologist visually inspected the axial T2w series and distinguished between the T2w and T2fs series. After this process was completed, 1,165 MRI series with a total of 8 classes were obtained. All series in the DLDS dataset covered the abdomen region. These series data were not used during training and were solely used as an external test dataset to evaluate the classification model.

\subsubsection{Clinical Proteomic Tumor Analysis Consortium Uterine Corpus Endometrial Carcinoma (CPTAC-UCEC) dataset}
The publicly available CPTAC-UCEC dataset \cite{13_cptac} contained 31 subjects with 902 pelvis MRI series from the National Cancer Institute’s CPTAC-UCEC cohort. There were many different series types in this dataset, such as coronal and sagittal sequences, which were excluded as they were not the axial sequences of primary interest to be studied in this paper. As a result, a total of 63 T2w and T2fs series were collected. This dataset covered the pelvis region and was only used as a test dataset for external validation.

\begin{table*}[!t]
\centering
\caption{Results of the comparison of classification models with different networks in Task 1}
\label{tab:result_task1}
\begin{tabular}{c|c|cccccc}
\hline
\textbf{\begin{tabular}[c]{@{}c@{}}Model \\ (Body part)\end{tabular}}        & \multicolumn{1}{c|}{\textbf{No. Studies}} & \multicolumn{1}{c}{\textbf{Precision}}                              & \multicolumn{1}{c}{\textbf{Sensitivity}}                            & \multicolumn{1}{c}{\textbf{Specificity}}                            & \multicolumn{1}{c}{\textbf{F1-score}}                               & \multicolumn{1}{c}{\textbf{Accuracy}}                               & \multicolumn{1}{c}{\textbf{AUC}}                                    \\ \hline
\multicolumn{8}{l}{\textbf{Test data covering all chest, abdomen, and pelvis region}}     \\ \hline
\begin{tabular}[c]{@{}c@{}}ResNet-50 \\ (All)\end{tabular}                   & 313                                      & \begin{tabular}[c]{@{}c@{}}0.959 \\ {[}0.952, 0.966{]}\end{tabular} & \begin{tabular}[c]{@{}c@{}}0.959 \\ {[}0.952, 0.966{]}\end{tabular} & \begin{tabular}[c]{@{}c@{}}0.995 \\ {[}0.993, 0.998{]}\end{tabular} & \begin{tabular}[c]{@{}c@{}}0.959 \\ {[}0.952, 0.966{]}\end{tabular} & \begin{tabular}[c]{@{}c@{}}0.966 \\ {[}0.960, 0.973{]}\end{tabular} & \begin{tabular}[c]{@{}c@{}}0.977 \\ {[}0.972, 0.982{]}\end{tabular} \\
\begin{tabular}[c]{@{}c@{}}ResNet-101 \\ (All)\end{tabular}                  & 313                                      & \begin{tabular}[c]{@{}c@{}}0.953 \\ {[}0.946, 0.961{]}\end{tabular} & \begin{tabular}[c]{@{}c@{}}0.953 \\ {[}0.945, 0.960{]}\end{tabular} & \begin{tabular}[c]{@{}c@{}}0.995 \\ {[}0.992, 0.997{]}\end{tabular} & \begin{tabular}[c]{@{}c@{}}0.953 \\ {[}0.946, 0.960{]}\end{tabular} & \begin{tabular}[c]{@{}c@{}}0.962 \\ {[}0.956, 0.969{]}\end{tabular} & \begin{tabular}[c]{@{}c@{}}0.974 \\ {[}0.968, 0.979{]}\end{tabular} \\
\begin{tabular}[c]{@{}c@{}}ResNet-152 \\ (All)\end{tabular}                  & 313                                      & \begin{tabular}[c]{@{}c@{}}0.953 \\ {[}0.945, 0.960{]}\end{tabular} & \begin{tabular}[c]{@{}c@{}}0.953 \\ {[}0.945, 0.960{]}\end{tabular} & \begin{tabular}[c]{@{}c@{}}0.995 \\ {[}0.992, 0.997{]}\end{tabular} & \begin{tabular}[c]{@{}c@{}}0.953 \\ {[}0.945, 0.960{]}\end{tabular} & \begin{tabular}[c]{@{}c@{}}0.962 \\ {[}0.955, 0.969{]}\end{tabular} & \begin{tabular}[c]{@{}c@{}}0.974 \\ {[}0.968, 0.979{]}\end{tabular} \\
\begin{tabular}[c]{@{}c@{}}EfficientNet-BN0 \\ (All)\end{tabular}            & 313                                      & \begin{tabular}[c]{@{}c@{}}0.951 \\ {[}0.943, 0.958{]}\end{tabular} & \begin{tabular}[c]{@{}c@{}}0.951 \\ {[}0.944, 0.959{]}\end{tabular} & \begin{tabular}[c]{@{}c@{}}0.995 \\ {[}0.992, 0.997{]}\end{tabular} & \begin{tabular}[c]{@{}c@{}}0.951 \\ {[}0.943, 0.958{]}\end{tabular} & \begin{tabular}[c]{@{}c@{}}0.961 \\ {[}0.954, 0.967{]}\end{tabular} & \begin{tabular}[c]{@{}c@{}}0.973 \\ {[}0.967, 0.979{]}\end{tabular} \\
\begin{tabular}[c]{@{}c@{}}EfficientNet-BN7 \\ (All)\end{tabular}            & 313                                      & \begin{tabular}[c]{@{}c@{}}0.957 \\ {[}0.950, 0.964{]}\end{tabular} & \begin{tabular}[c]{@{}c@{}}0.957 \\ {[}0.950, 0.964{]}\end{tabular} & \begin{tabular}[c]{@{}c@{}}0.995 \\ {[}0.993, 0.998{]}\end{tabular} & \begin{tabular}[c]{@{}c@{}}0.957 \\ {[}0.950, 0.964{]}\end{tabular} & \begin{tabular}[c]{@{}c@{}}0.965 \\ {[}0.959, 0.972{]}\end{tabular} & \begin{tabular}[c]{@{}c@{}}0.976 \\ {[}0.971, 0.981{]}\end{tabular} \\
\begin{tabular}[c]{@{}c@{}}DenseNet-121 \\ (All)\end{tabular}                & 313                                      & \begin{tabular}[c]{@{}c@{}}0.966 \\ {[}0.960, 0.972{]}\end{tabular}  & \begin{tabular}[c]{@{}c@{}}0.966 \\ {[}0.960, 0.972{]}\end{tabular}  & \begin{tabular}[c]{@{}c@{}}0.996 \\ {[}0.994, 0.998{]}\end{tabular}  & \begin{tabular}[c]{@{}c@{}}0.966 \\ {[}0.960, 0.972{]}\end{tabular}  & \begin{tabular}[c]{@{}c@{}}0.972 \\ {[}0.967, 0.978{]}\end{tabular}  & \begin{tabular}[c]{@{}c@{}}0.981 \\ {[}0.976, 0.986{]}\end{tabular}  \\ \hline

\multicolumn{8}{l}{\textbf{Test data covering each body region}}   \\ \hline
\begin{tabular}[c]{@{}c@{}}DenseNet-121 \\ (Chest)\end{tabular}              & 4                                        & \begin{tabular}[c]{@{}c@{}}0.958 \\ {[}0.896, 1.000{]}\end{tabular} & \begin{tabular}[c]{@{}c@{}}0.938 \\ {[}0.862, 1.000{]}\end{tabular} & \begin{tabular}[c]{@{}c@{}}0.993 \\ {[}0.967, 1.000{]}\end{tabular} & \begin{tabular}[c]{@{}c@{}}0.933 \\ {[}0.856, 1.000{]}\end{tabular} & \begin{tabular}[c]{@{}c@{}}0.950 \\ {[}0.882, 1.000{]}\end{tabular} & \begin{tabular}[c]{@{}c@{}}0.965 \\ {[}0.909, 1.000{]}\end{tabular} \\
\begin{tabular}[c]{@{}c@{}}DenseNet-121 \\ (Abdomen)\end{tabular}            & 292                                      & \begin{tabular}[c]{@{}c@{}}0.966 \\ {[}0.960, 0.973{]}\end{tabular} & \begin{tabular}[c]{@{}c@{}}0.966 \\ {[}0.960, 0.973{]}\end{tabular} & \begin{tabular}[c]{@{}c@{}}0.996 \\ {[}0.994, 0.998{]}\end{tabular} & \begin{tabular}[c]{@{}c@{}}0.966 \\ {[}0.960, 0.973{]}\end{tabular} & \begin{tabular}[c]{@{}c@{}}0.973 \\ {[}0.967, 0.979{]}\end{tabular} & \begin{tabular}[c]{@{}c@{}}0.981 \\ {[}0.976, 0.986{]}\end{tabular} \\
\begin{tabular}[c]{@{}c@{}}DenseNet-121 \\ (Abdomen+Pelvis)\end{tabular} & 8                                        & \begin{tabular}[c]{@{}c@{}}0.970 \\ {[}0.933, 1.000{]}\end{tabular} & \begin{tabular}[c]{@{}c@{}}0.969 \\ {[}0.931, 1.000{]}\end{tabular} & \begin{tabular}[c]{@{}c@{}}0.997 \\ {[}0.984, 1.000{]}\end{tabular} & \begin{tabular}[c]{@{}c@{}}0.969 \\ {[}0.931, 1.000{]}\end{tabular} & \begin{tabular}[c]{@{}c@{}}0.975 \\ {[}0.941, 1.000{]}\end{tabular} & \begin{tabular}[c]{@{}c@{}}0.983 \\ {[}0.954, 1.000{]}\end{tabular} \\
\begin{tabular}[c]{@{}c@{}}DenseNet-121 \\ (Pelvis)\end{tabular}             & 9                                        & \begin{tabular}[c]{@{}c@{}}0.960 \\ {[}0.919, 1.000{]}\end{tabular} & \begin{tabular}[c]{@{}c@{}}0.958 \\ {[}0.917, 1.000{]}\end{tabular} & \begin{tabular}[c]{@{}c@{}}0.995 \\ {[}0.981, 1.000{]}\end{tabular} & \begin{tabular}[c]{@{}c@{}}0.958 \\ {[}0.917, 1.000{]}\end{tabular} & \begin{tabular}[c]{@{}c@{}}0.967 \\ {[}0.930, 1.000{]}\end{tabular} & \begin{tabular}[c]{@{}c@{}}0.977 \\ {[}0.946, 1.000{]}\end{tabular} \\
\hline
\multicolumn{8}{l}{
\textit{Note.} The 95\% confidence interval is in each bracket. }
\end{tabular}
\end{table*}

\subsection{Body MRI Series Classification with Deep Learning}
An overview of the study design is shown in Fig.~\ref{fig: pipeline}(A). Our classification model takes a 3D MRI series as input, estimates the probability score for each sequence, and predicts the class corresponding to the input. We conducted the following four tasks to identify the best-performing network amongst the existing models and evaluated the performance of the best classifier. Fig.~\ref{fig: pipeline}(B) provides detailed descriptions of the data splits and networks used for training and testing in each task. 

\subsubsection{Task 1 - Comparing Classifiers}
We investigated the performance of several 3D networks including ResNet-50, ResNet-101, ResNet-152 \cite{17_he2016deep}, EfficientNet-BN0, EfficientNet-BN7 \cite{18_tan2019efficientnet}, and DenseNet-121 \cite{19_huang2017densely}. These networks were originally developed for classifying natural images and have shown impressive classification performance. Each network was trained from scratch with 1,363 studies from the Siemens scanners in the NIH dataset, in which the studies were split into training and validation sets at the patient level with an 88:12 ratio, and tested on 313 studies from the same manufacturer in the internal NIH dataset. 

\subsubsection{Task 2 - Evaluation of training data quantity}
We assessed the susceptibility of the best-performing model in Task 1 according to the quantity of data used for training. We trained the model using {20\%, 40\%, 60\%, 80\%} of the data (n=1,363 studies) acquired with Siemens scanners in the NIH dataset. For each model with different amounts of training data, the data was split into training and validation at a ratio of 88:12 for cross-validation. The model performance was then evaluated on the 313 test studies obtained by the same scanners in the internal NIH dataset. 

\subsubsection{Task 3 - Evaluation of robustness on out-of-training-data distribution} 
We evaluated the robustness of the best-performing classifier in Task 1 on the out-of-training distribution data. To determine the generalization capacity of the trained model from Task 1, its performance was measured on the following external datasets not included in the training set: the Philips and GE scanner studies in the NIH dataset and the external DLDS and CPTAC-UCEC datasets. 

\subsubsection{Task 4 - Evaluation of using different scanner data in model training}
To study how the model performs when training with data from different MRI scanners, we trained the model in two strategies: one is to train the model from scratch by combining the two scanner datasets (Strategy 1), and the other is to fine-tune the pre-trained model using a different scanner dataset (Strategy 2). For strategy 1, an equal number of studies (n=82) from both the Siemens and Philips scanners in the NIH dataset were used to train the classification model. Then, the model performance was tested on the remaining Siemens studies (n=1,594) and Philips studies (n=81) of the internal NIH dataset. For strategy 2, we finetuned the model pre-trained in Task 1 using the Philips scanner studies (n=82) and evaluated the model on the remaining Siemens studies (n=313) and Philips studies (n=81) in the NIH dataset. Here, we used 20\% of the data in the training set as validation data for cross-validation.

\subsection{Model Training}

In the training stage, a classification network $G$ took a 3D MRI volume $x$ and estimated the class probabilities:
\begin{align}
    \mathrm{output}=P(\hat{y}=y|G(x)),
\end{align}
where $G(x)$ is the output of the classifier, $\hat{y}$ is the predicted sequence type, and $y$ is the ground-truth label corresponding to the input. For the objective function, we employed a cross-entropy loss that can be computed by:
\begin{align}
     \mathcal{L}(y, \hat{y}) = -\sum_i y_i \mathrm{log}(\hat{y}_i).
\end{align}
During training, the model weights associated with the epoch that achieved the highest validation accuracy were saved. In the inference stage, the classification model with fixed parameters $G^*$ was tested by predicting the series type for an input volume via the following argmax function:
\begin{align}
    \mathrm{argmax}_y P(\hat{y}=y|G(x)). 
\end{align}
As we conducted five-fold cross-validation in our experiments, we ensembled the model output by averaging the probability maps and obtained the final prediction for the input volume. 

\section{Experiments and Results}

\begin{figure*}[!t]
\centering
\centerline{\includegraphics[width=\linewidth]{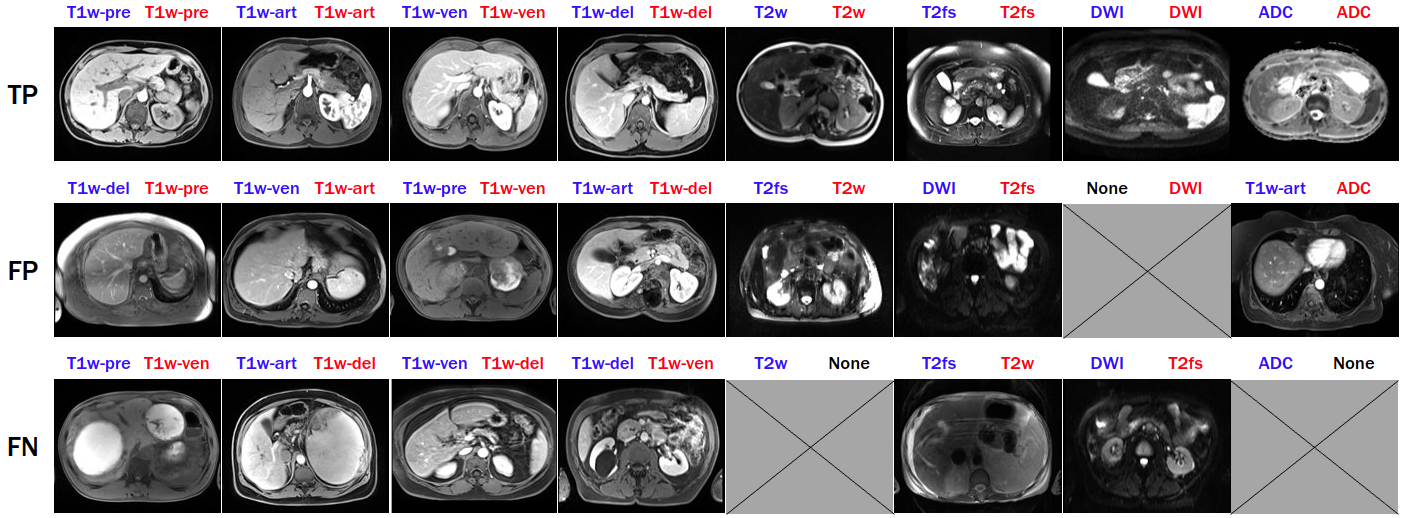}}
\caption{Example images of true positive (TP), false positive (FP), and false negative (FN) cases from the DenseNet-121 model for body MRI classification with the NIH dataset in Task 1. The images come from different patients. For each image, the blue color denotes the ground-truth label, and the red color denotes the prediction. None refers that there is no corresponding case.}
\label{fig: result_task1}
\end{figure*}

\subsection{Data Processing}
Since the MRI series volume from different protocols, scanners, and sources has different spatial dimensions, we processed the MRI data before feeding it into the models. Specifically, we first resampled each 3D MRI data to have the voxel size of $1.5\times1.5\times7.8 mm$ by a Nibabel library \cite{brett_2023_7795644} in Python, \add{where the voxel size was set to the median value of each spatial dimension of MRI volumes in the NIH dataset to have a consistent voxel size for the input obtained from different acquisition parameters and scanners.} Then, using a library of the Medical Open Network for Artificial Intelligence (MONAI) \cite{22_cardoso2022monai}, we clipped the volume intensities to the 1st and 99th percentile \cite{20_kociolek2020does} \add{to enhance image contrast} and resized the volume to a spatial size of $256\times256\times36$ by either cropping or padding the volume. Here, if the volume spatial dimension was larger than the target size, we did central cropping, otherwise, we did symmetric padding along the dimension. Lastly, we changed the volume orientation to (right, left), (posterior, anterior), and (inferior, superior) for the axis directions. Through this data preprocessing, we obtained the input data and implemented the classification models.

\subsection{Implementation details}
For our 3D MRI series classification model, we built the 3D networks using the MONAI library \cite{22_cardoso2022monai} in Python. While the networks of ResNet, EfficientNet, and DenseNet were originally designed for 2D images, by changing the 2D layers to 3D layers, such as 2D convolutional layers to 3D convolutional layers, the networks can be readily structured for 3D volumes. The MONAI library provides the 3D network architectures in this way, and by using the library, we implemented the 3D networks and updated the network parameters from scratch using our 3D MRI data.

For training the model, we performed five-fold cross-validation. In each fold, the batch size was set to 2, and the model was trained for 25 epochs through an Adam optimization \cite{21_kingma2014adam} with a learning rate of 0.0001. At the test stage, the model with the highest validation accuracy in each fold was used. The final prediction was calculated as the class with the highest probability from the average of all 5-fold probability maps. All experiments were implemented using PyTorch in Python with a single Nvidia A100-SXM4-40GB graphics processing unit (GPU). The source code of our work is available at \url{https://github.com/boahK/MRI_Classifier}. 

\subsection{Evaluation Metrics and Statistical Analysis}
To quantify the performance of classification models, the following metrics were computed: precision, sensitivity, specificity, F1-score, accuracy, and area under the receiver operating characteristic curve (AUC). These metrics included the 95\% confidence interval (C.I.). Also, to analyze the performance difference between models, McNemar’s test \cite{23_mcnemar1947note} was conducted for Tasks 1 and 4 to compare their predictive accuracy. A p-value$<$0.05 was considered to be statistically significant.

\begin{figure}[!t]
\centering
\centerline{\includegraphics[width=\linewidth]{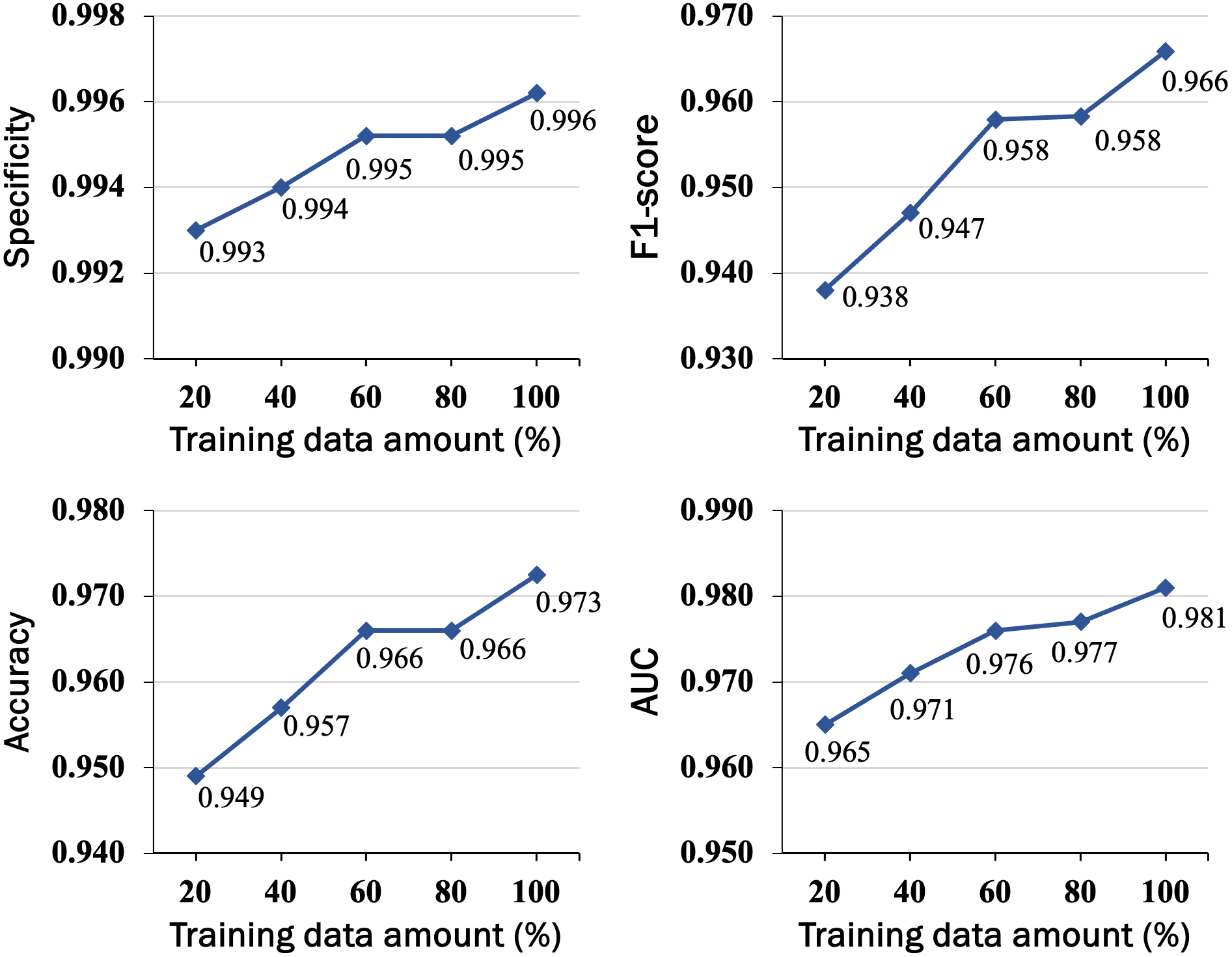}}
\caption{Classification performance of the DenseNet-121 model according to the amount of data used for training the model as described in Task 2. The specificity, F1-score, accuracy, and area under the receiver operating characteristic curve (AUC) against the training data amount are visualized, where the F1-score was computed as the harmonic mean of the precision and recall.}
\label{fig: result_task2}
\end{figure}

\begin{table*}[!t]
\centering
\caption{Results of testing on the external out-of-data distribution datasets in Task 3 }
\label{tab:result_task3}
\begin{tabular}{c|cccccc}
\hline
\multicolumn{1}{c|}{\textbf{Test data}}                              & \multicolumn{1}{c}{\textbf{Precision}}                               & \multicolumn{1}{c}{\textbf{Sensitivity}}                             & \multicolumn{1}{c}{\textbf{Specificity}}                             & \multicolumn{1}{c}{\textbf{F1-score}}                                & \multicolumn{1}{c}{\textbf{Accuracy}}                                & \multicolumn{1}{c}{\textbf{AUC}}                                     \\ \hline
\multicolumn{7}{l}{\textbf{Baseline: Internal test data}}  \\ \hline
[NIH] Siemens dataset  & \begin{tabular}[c]{@{}c@{}}0.966 \\ {[}0.960, 0.972{]}\end{tabular}  & \begin{tabular}[c]{@{}c@{}}0.966 \\ {[}0.960, 0.972{]}\end{tabular}  & \begin{tabular}[c]{@{}c@{}}0.996 \\ {[}0.994, 0.998{]}\end{tabular}  & \begin{tabular}[c]{@{}c@{}}0.966 \\ {[}0.960, 0.972{]}\end{tabular}  & \begin{tabular}[c]{@{}c@{}}0.972 \\ {[}0.967, 0.978{]}\end{tabular}  & \begin{tabular}[c]{@{}c@{}}0.981 \\ {[}0.976, 0.986{]}\end{tabular}  \\ \hline

\multicolumn{7}{l}{\textbf{Comparisons: External test data}}  \\ \hline
[NIH] Philips dataset & \begin{tabular}[c]{@{}c@{}}0.887  \\ {[}0.871, 0.903{]}\end{tabular} & \begin{tabular}[c]{@{}c@{}}0.875  \\ {[}0.858, 0.891{]}\end{tabular} & \begin{tabular}[c]{@{}c@{}}0.984  \\ {[}0.978, 0.991{]}\end{tabular} & \begin{tabular}[c]{@{}c@{}}0.875  \\ {[}0.858, 0.891{]}\end{tabular} & \begin{tabular}[c]{@{}c@{}}0.889  \\ {[}0.873, 0.905{]}\end{tabular} & \begin{tabular}[c]{@{}c@{}}0.930  \\ {[}0.916, 0.943{]}\end{tabular} \\

[NIH] GE dataset & \begin{tabular}[c]{@{}c@{}}0.788  \\ {[}0.722, 0.854{]}\end{tabular} & \begin{tabular}[c]{@{}c@{}}0.752  \\ {[}0.681, 0.822{]}\end{tabular} & \begin{tabular}[c]{@{}c@{}}0.978  \\ {[}0.954, 1.000{]}\end{tabular} & \begin{tabular}[c]{@{}c@{}}0.759  \\ {[}0.690, 0.829{]}\end{tabular} & \begin{tabular}[c]{@{}c@{}}0.849  \\ {[}0.791, 0.907{]}\end{tabular} & \begin{tabular}[c]{@{}c@{}}0.865  \\ {[}0.809, 0.920{]}\end{tabular} \\ 

[DLDS] Abdomen dataset & \begin{tabular}[c]{@{}c@{}}0.827  \\ {[}0.805, 0.849{]}\end{tabular} & \begin{tabular}[c]{@{}c@{}}0.857  \\ {[}0.837, 0.877{]}\end{tabular} & \begin{tabular}[c]{@{}c@{}}0.981  \\ {[}0.973, 0.989{]}\end{tabular} & \begin{tabular}[c]{@{}c@{}}0.835  \\ {[}0.814, 0.856{]}\end{tabular} & \begin{tabular}[c]{@{}c@{}}0.872  \\ {[}0.853, 0.891{]}\end{tabular} & \begin{tabular}[c]{@{}c@{}}0.919  \\ {[}0.903, 0.935{]}\end{tabular} \\ 

[CPTAC-UCEC] Pelvis dataset  & \begin{tabular}[c]{@{}c@{}}1.000 \\ {[}1.000, 1.000{]}\end{tabular}  & \begin{tabular}[c]{@{}c@{}}0.812 \\ {[}0.716, 0.909{]}\end{tabular}  & \begin{tabular}[c]{@{}c@{}}1.000  \\ {[}1.000, 1.000{]}\end{tabular} & \begin{tabular}[c]{@{}c@{}}0.885 \\ {[}0.806, 0.964{]}\end{tabular}  & \begin{tabular}[c]{@{}c@{}}0.810 \\ {[}0.713, 0.906{]}\end{tabular}  & \begin{tabular}[c]{@{}c@{}}0.906 \\ {[}0.834, 0.978{]}\end{tabular} \\ \hline
\multicolumn{7}{l}{
\textit{Note.} The 95\% confidence interval is in each bracket. }
\end{tabular}
\end{table*}

\subsection{Results}
\subsubsection{Task 1 - Comparing classifiers}
Table~\ref{tab:result_task1} lists the performance of the different classification models tested in Task 1. The DenseNet-121 model showed the best performance across all evaluation metrics with a precision of 0.966, an F1-score of 0.966, and an accuracy of 0.972. There was a statistically significant difference (p=0.02) in comparison to the ResNet-50 network, which posted the second-highest accuracy. The EfficientNet class of models had the lowest values for all metrics compared to the ResNet and DenseNet models. Also, when evaluating the DenseNet-121 model for each body region, the model achieved an accuracy of 0.950, 0.973, and 0.967 for the chest, abdomen, and pelvis region, respectively. This represented that the performance for each body region was similar to that computed on all test data when considering the number of studies of each body region. Based on these results, the DenseNet-121 network was used as the base classification network for the subsequent Tasks 2-4. 

Example images of true positive (TN), false positive (FP), and false negative (FN) predictions from DenseNet-121 are shown in Fig.~\ref{fig: result_task1}. The misclassification of FP and FN occurred when the MRI series had similar voxel intensity patterns. As the contrasts of the T1-weighted series varied due to patient and scanner differences, the majority of the FP and FNs appeared when classifying the T1-weighted series. Also, when the voxel intensities of T2fs and DWI were low, the series types were classified incorrectly as they appeared to be similar. Compared to these cases, when the MRI series types had distinguishable features, such as T2w and ADC series, the model did not misclassify these series types. \add{The further erroneous case analysis with the confusion matrix can be found in Section~\ref{section:result_task3}.}

\begin{figure}[!t]
\centering
\centerline{\includegraphics[width=\linewidth]{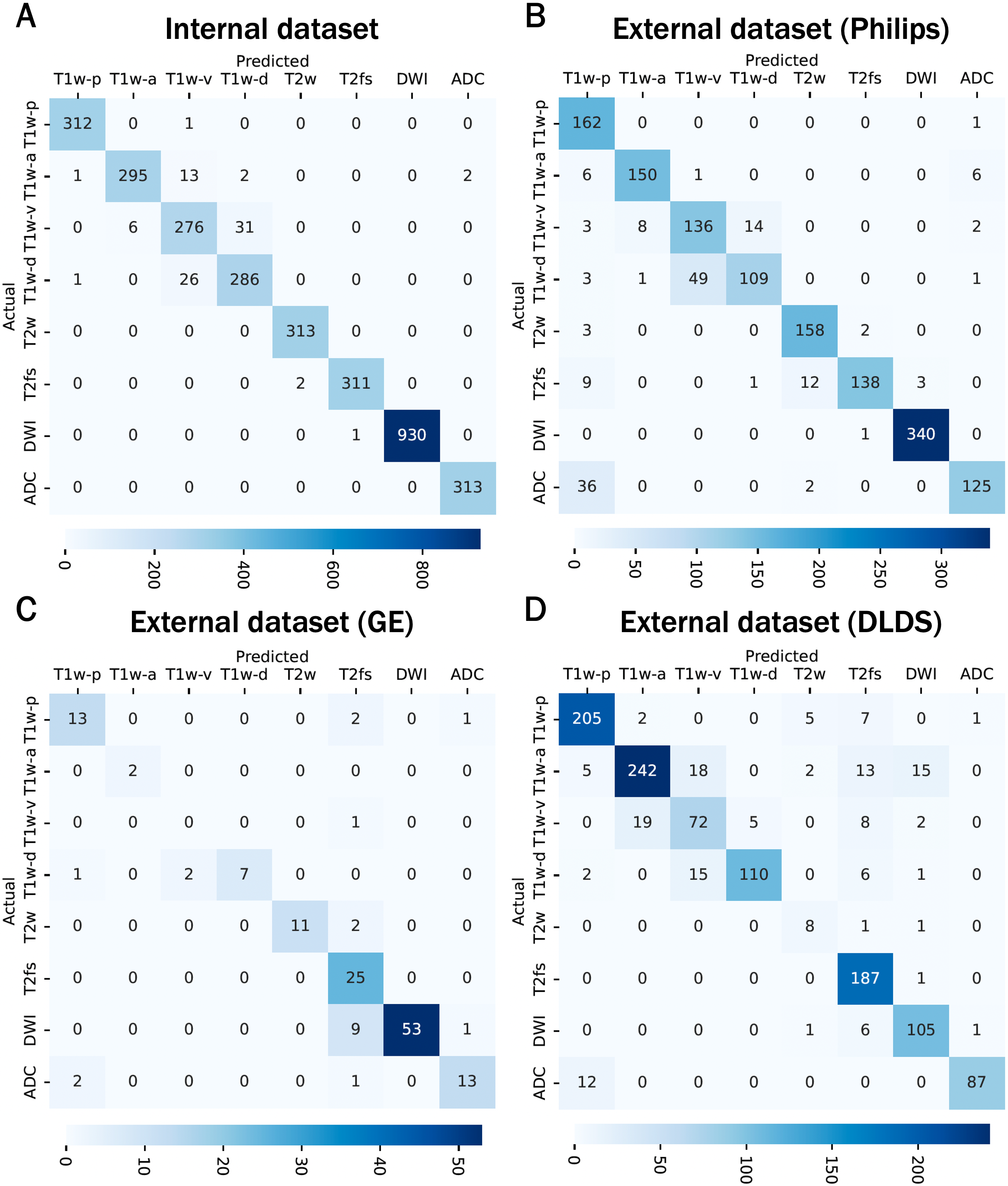}}
\caption{Confusion matrices of the predictions on various test datasets for the model trained with the NIH Siemens dataset in Task 3. T1w-p, T1w-a, T1w-v, and T1-d denote pre-contrast, arterial, portal venous, and delayed T1-weighted imaging, respectively. The blue bar located at the bottom of each matrix indicates the number of MRI series. (A) Confusion matrix on the internal NIH Siemens test dataset. (B) Confusion matrix on the NIH Philips test dataset. (C) Confusion matrix on the NIH GE test dataset. (D) Confusion matrix on the external DLDS dataset.}
\label{fig: result_task3}
\end{figure}

\subsubsection{Task 2 - Effect of training data quantities  }
Fig.~\ref{fig: result_task2} shows the DenseNet-121 model performance when trained with varying training data quantities. With 20\% of the training data, the specificity, F1-score, and accuracy of the model were 0.993, 0.938, and 0.949, respectively. The values of these metrics increased with larger amounts of training data, and models using 60\% and 80\% of training data achieved similar performance. In particular, the model using 100\% of the training data performed best with an accuracy of 0.972, but models using more than 60\% of the training data also achieved an accuracy of 0.966 or higher. The AUC also improved from 0.965 to 0.981 as the training data quantity increased. These results indicated that 60\% of training data was the plateau point, and 729 studies were sufficient to train the DenseNet-121 model for MRI series classification with $\geq$96\% accuracy.

\begin{table*}[!t]
\centering
\caption{Results on the classification models trained with different domain data in Task 4 }
\label{tab:result_task4}
\begin{tabular}{c|c|cccccc}
\hline
{\textbf{Training data}}  & \textbf{Test data}  & \textbf{Precision}   & \textbf{Sensitivity}   & \textbf{Specificity}  & \textbf{F1-score}    & \textbf{Accuracy}    & \textbf{AUC}  \\ \hline
\multicolumn{8}{l}{\textbf{Baseline}}   \\ \hline
\multirow{3}{*}{Siemens (n=1,363)}   & Siemens (n=313)  & \begin{tabular}[c]{@{}c@{}}0.966 \\ {[}0.960, 0.972{]}\end{tabular}  & \begin{tabular}[c]{@{}c@{}}0.966 \\ {[}0.960, 0.972{]}\end{tabular}  & \begin{tabular}[c]{@{}c@{}}0.996 \\ {[}0.994, 0.998{]}\end{tabular}  & \begin{tabular}[c]{@{}c@{}}0.966 \\ {[}0.960, 0.972{]}\end{tabular}  & \begin{tabular}[c]{@{}c@{}}0.972 \\ {[}0.967, 0.978{]}\end{tabular}  & \begin{tabular}[c]{@{}c@{}}0.981 \\ {[}0.976, 0.986{]}\end{tabular}  \\
 & Philips (n=81) & \begin{tabular}[c]{@{}c@{}}0.883  \\ {[}0.860, 0.906{]}\end{tabular} & \begin{tabular}[c]{@{}c@{}}0.867  \\ {[}0.842, 0.891{]}\end{tabular} & \begin{tabular}[c]{@{}c@{}}0.984  \\ {[}0.975, 0.993{]}\end{tabular} & \begin{tabular}[c]{@{}c@{}}0.867  \\ {[}0.842, 0.891{]}\end{tabular} & \begin{tabular}[c]{@{}c@{}}0.884  \\ {[}0.861, 0.907{]}\end{tabular} & \begin{tabular}[c]{@{}c@{}}0.925  \\ {[}0.906, 0.944{]}\end{tabular} \\ \hline
       
\multicolumn{8}{l}{\textbf{Strategy 1: Training from scratch}}   \\ \hline
\begin{tabular}[c]{@{}c@{}} \\ Siemens (n=82) \end{tabular} & Siemens (n=1,594) & \begin{tabular}[c]{@{}c@{}}0.919  \\ {[}0.915, 0.923{]}\end{tabular} & \begin{tabular}[c]{@{}c@{}}0.916  \\ {[}0.912, 0.921{]}\end{tabular} & \begin{tabular}[c]{@{}c@{}}0.991  \\ {[}0.989, 0.992{]}\end{tabular} & \begin{tabular}[c]{@{}c@{}}0.915  \\ {[}0.911, 0.919{]}\end{tabular} & \begin{tabular}[c]{@{}c@{}}0.933  \\ {[}0.929, 0.937{]}\end{tabular} & \begin{tabular}[c]{@{}c@{}}0.954  \\ {[}0.950, 0.957{]}\end{tabular} \\ 
\begin{tabular}[c]{@{}c@{}} + Philips (n=82) \\ \, \end{tabular} & Philips (n=81) & \begin{tabular}[c]{@{}c@{}}0.917  \\ {[}0.897, 0.937{]}\end{tabular} & \begin{tabular}[c]{@{}c@{}}0.912  \\ {[}0.892, 0.932{]}\end{tabular} & \begin{tabular}[c]{@{}c@{}}0.989  \\ {[}0.982, 0.997{]}\end{tabular} & \begin{tabular}[c]{@{}c@{}}0.909  \\ {[}0.889, 0.930{]}\end{tabular} & \begin{tabular}[c]{@{}c@{}}0.924  \\ {[}0.905, 0.943{]}\end{tabular} & \begin{tabular}[c]{@{}c@{}}0.951  \\ {[}0.935, 0.966{]}\end{tabular} \\
 \hline

 \multicolumn{8}{l}{\textbf{Strategy 2: Finetuning pre-trained model}}   \\ \hline
\begin{tabular}[c]{@{}c@{}} \\ Siemens (n=1,363) \end{tabular} & {Siemens (n=313)} & \begin{tabular}[c]{@{}c@{}}{0.954}  \\ {[0.946, 0.961]}\end{tabular} & \begin{tabular}[c]{@{}c@{}}{0.952}  \\ {[0.944, 0.959]}\end{tabular} & \begin{tabular}[c]{@{}c@{}}{0.995}  \\ {[0.992, 0.997]}\end{tabular} & \begin{tabular}[c]{@{}c@{}}{0.952}  \\ {[0.944, 0.959]}\end{tabular} & \begin{tabular}[c]{@{}c@{}}{0.961}  \\ {[0.954, 0.968]}\end{tabular} & \begin{tabular}[c]{@{}c@{}}{0.973}  \\ {[0.967, 0.979]}\end{tabular} \\
\begin{tabular}[c]{@{}c@{}} + Philips (n=82) \\ \, \end{tabular} & Philips (n=81) & \begin{tabular}[c]{@{}c@{}}0.946  \\ {[}0.930, 0.962{]}\end{tabular} & \begin{tabular}[c]{@{}c@{}}0.946  \\ {[}0.930, 0.962{]}\end{tabular} & \begin{tabular}[c]{@{}c@{}}0.993  \\ {[}0.988, 0.999{]}\end{tabular} & \begin{tabular}[c]{@{}c@{}}0.946  \\ {[}0.930, 0.962{]}\end{tabular} & \begin{tabular}[c]{@{}c@{}}0.953  \\ {[}0.938, 0.968{]}\end{tabular} & \begin{tabular}[c]{@{}c@{}}0.970  \\ {[}0.957, 0.982{]}\end{tabular} \\ \hline

\multicolumn{8}{l}{
\textit{Note.} The 95\% confidence interval is in each bracket. }
\end{tabular}
\end{table*}

\subsubsection{Task 3 - Robustness to external datasets}\label{section:result_task3}
To evaluate the robustness of the DenseNet-121 model from Task 1 (trained with the Siemens scanner data from the NIH dataset) to external datasets, we tested it on the Philips scanner studies and the GE scanner studies from the NIH dataset, on the external DLDS dataset, and on the external CPTAC-UCEC dataset. Table~\ref{tab:result_task3} reports the results of this experiment. When comparing the classification performance on the different scanner datasets to that on the internal Siemens dataset, the F1-score and accuracy on the Philips dataset were 0.875 and 0.889, respectively, which decreased by about 9\%. Also, the F1-score and accuracy on the GE dataset were 0.759 and 0.849, respectively.
On the other hand, when testing the model on the external datasets of different institutions, for the external DLDS dataset, the model obtained the F1-score and accuracy of 0.834 and 0.872 respectively. Also, for the CPTAC-UCEC dataset, the F1-score and accuracy of the model were 0.885 and 0.810.  

Fig.~\ref{fig: result_task3} shows the confusion matrix for each test dataset, where the matrix shows the number of predictions of each class versus the actual class. \add{Specifically, in the internal Siemens test dataset (Fig.~\ref{fig: result_task3}(A)), the most erroneous cases occurred in different dynamic contrast-enhanced T1-weighted series. In particular, the model incorrectly classified between the T1-weighted data in the venous phase and the data in the delayed phase, which might result from the inconsistent acquisition time and different blood flow velocities according to the patients. These incorrect predictions between the different T1-weighted data were similarly shown in the Philips test dataset (Fig.~\ref{fig: result_task3}(B)) and the external DLDS test dataset (Fig.~\ref{fig: result_task3}(D)). However, in the external datasets, we could observe additional misclassifications beyond the T1-weighted data. In the Philips test dataset (Fig.~\ref{fig: result_task3}(B)), the model misclassified the ADC series significantly as the pre-phase T1-weighted series. Also, in the GE dataset (Fig.~\ref{fig: result_task3}(C)), the predictions with fat-suppressed T2-weighted for the DWI series resulted in the most misclassifications. Moreover, in the DLDS dataset (Fig.~\ref{fig: result_task3}(D)), the model had trouble discriminating the arterial T1-weighted series from the fat-suppressed T2-weighted and DWI series data. These results may come from different scanner-specific MRI features and different imaging acquisition protocols of institutions, as well as the smaller number of data in the external datasets compared to that in the internal Siemens test dataset.} Fortunately, the degraded classification performance caused by scanner-specific features can be readily addressed by fine-tuning the model using small amounts of data from different scanners, as demonstrated in the next Section~\ref{sec:result_task4}.


\subsubsection{Task 4 - Model training using different scanner data}\label{sec:result_task4}
Table~\ref{tab:result_task4} shows the performance of the DenseNet-121 model trained with the two different strategies. Compared to the baseline model trained only using the Siemens dataset, the models from the two strategies showed similar precision, sensitivity, specificity, and AUC values between the Siemens and Philips test data. Also, on the Philips dataset, the models achieved significantly improved classification performance ($p<0.001$) over the baseline model. In particular, although the total number of training data in Strategy 1 was much lower than that in the baseline model, we could observe that the model in Strategy 1 achieved a 4\% gain in the F1-score and accuracy on the Philips dataset. In addition, when comparing the two strategies, the finetuned model in Strategy 2 further enhanced the performance of classifying the Philips data over the model in Strategy 1 while minimizing the performance degradation on the Siemens dataset. This result indicates that the model can handle data from various MRI scanners with high performance by simply finetuning the model using a small amount of different scanner data.

\section{Discussion}
In this work, a deep learning framework to classify 8 different series in an MRI exam was presented and tested on both internal and external datasets. Our results indicated that the DenseNet-121 model trained on internally acquired Siemens studies outperformed the ResNet and EfficientNet models. Using 60\% of the training data was sufficient to achieve an accuracy $\geq$0.96, but performance improved as the training data quantity increased. The DenseNet-121 model trained on only Siemens studies also performed well when evaluating it on the unseen NIH Philips and GE dataset, the DLDS dataset, and the CPTAC-UCEC dataset. When training the model with combined Siemens and Philips data, the accuracy on the Philips test dataset was higher than that of the model trained only with Siemens data ($p<0.001$). 

\subsection{Key Aspects of This Work}
To the best of our knowledge, there is no prior work for the task of classifying various body MRI series. In comparison to previous deep learning-based methods for MRI series classification \cite{5_zhu20203d, 6_baumgartner2023metadata}, our approach achieved 0.972 accuracy using data that includes chest, abdomen, and pelvis regions. While Zhu et al. \cite{5_zhu20203d} showed 0.846 accuracy for classifying 30 series types, their work handled only abdominal MRI studies even though the number of series was larger than those considered in our study. Moreover, when grouping certain series types that were regarded by a radiologist as less significant or non-significant for clinical practice, although the accuracy of 0.904 was similar to ours, the total amount of series for this metric was unknown. Additionally, Baumgartner et al. \cite{6_baumgartner2023metadata} presented a deep-learning-based classifier on prostate MRI exams. Their work achieved 0.999 accuracy for the classification of 10 different series, whereas our model showed only 2\% lower accuracy with 0.966 sensitivity and 0.996 specificity. Finally, our model was tested on external data including exams from different scanners (Philips and GE) and different institutions (DLDS and CPTAC-UCEC), which yielded an accuracy of $\geq$0.81.

\begin{figure}[!t]
\centering
\centerline{\includegraphics[width=\linewidth]{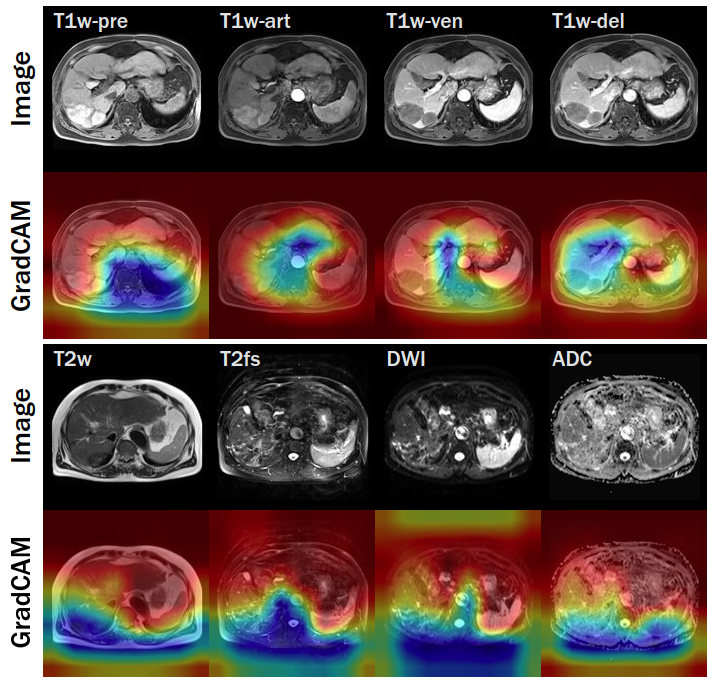}}
\caption{Visualization of the MRI series classification by our DenseNet-121 model for each series type with GradCAM \cite{selvaraju2017grad}. For each GradCAM image, the red areas are the most salient activation parts, while the blue areas are the least salient parts.}
\label{fig: discuss_gradcam}
\end{figure}

Another key aspect of this work was the capability to accurately classify contrast-enhanced T1-weighted series where the contrast varied according to the time after the contrast agent injection \cite{24_jackson2005dynamic}. Although these T1-weighted series have different contrasts, they are difficult to distinguish due to inconsistent data acquisition times post the injection of contrast agents and different blood flow velocities in each patient. Also, as shown in Fig.~\ref{fig: discuss_gradcam}, when comparing T1w-ven and T1-del series images, the bright regions are very slightly different from each other. Nevertheless, our model learned distinct image features for each series type and classified those challenging cases with high performance – The F1-score for each T1w-art, T1w-ven, and T1-del was 0.961, 0.878, and 0.905, respectively. Here, the lower performance on the T1w-por and T1w-del may be due to different data scanning times for each MRI exam, which is further exacerbated when exams are obtained across different scanners and institutions.

On the other hand, for the other MRI series types than the contrast-enhanced T1-weighted series, each series has differences mainly in the overall image contrast and intensities, and there are also specific distinct features for each series, which enables the network to classify the series type. Specifically, as shown in Fig.~\ref{fig: discuss_gradcam}, when comparing T1w and T2w series images, the fluid-based tissue in T1w is dark but that in T2w is bright, and the fat of the body in T2w is brighter than the other organs. Also, T2fs, DWI, and ADC show the spine brighter than the other regions, where these three series have different image features and intensity ranges.

To show that the network in our work determined the MRI series types according to the different features of each series, we visualized the activation map of the network for each series through GradCAM \cite{selvaraju2017grad} using \cite{jacobgilpytorchcam}. From the GradCAM images in Fig.~\ref{fig: discuss_gradcam}, we could observe that for the T1w, T2, T2fs, DWI, and ADC data, the network performed the series classification mainly based on the overall abdominal organs and tissues other than the spine and the adjacent regions. On the other hand, the network classified contrast-enhanced T1-weighted series of T1w-art, T1w-ven, and T1w-del based on the different salient regions around the aorta, the inferior vena cava, and the portal vein in the activation map. These indicate that the capability of the network to extract different features for each series type allows our model to distinguish various types of MRI series including the challenging dynamic contrast series.

\begin{table}[!t]
\centering
\caption{Results of the study on performance stability of our DenseNet-121 classification model}
\label{tab:discuss_seeds}
\begin{tabular}{c|ccc}
\hline
\multicolumn{1}{c|}{\textbf{Seed}} & \multicolumn{1}{c}{\textbf{Specificity}}                             & \multicolumn{1}{c}{\textbf{F1-score}}                                & \multicolumn{1}{c}{\textbf{Accuracy}}       \\ \hline
1 & \begin{tabular}[c]{@{}c@{}}0.996 \\ {[}0.994, 0.998{]}\end{tabular}  & \begin{tabular}[c]{@{}c@{}}0.966 \\ {[}0.960, 0.972{]}\end{tabular}  & \begin{tabular}[c]{@{}c@{}}0.972 \\ {[}0.967, 0.978{]}\end{tabular}  \\
{2} & \begin{tabular}[c]{@{}c@{}}{0.995}  \\ {[0.993, 0.998]}\end{tabular} & \begin{tabular}[c]{@{}c@{}}{0.957}  \\ {[0.950, 0.964]}\end{tabular} & \begin{tabular}[c]{@{}c@{}}{0.965}  \\ {[0.959, 0.972]}\end{tabular} \\ 
{3} & \begin{tabular}[c]{@{}c@{}}{0.996}  \\ {[0.993, 0.998]}\end{tabular} & \begin{tabular}[c]{@{}c@{}}{0.960}  \\ {[0.953, 0.967]}\end{tabular} & \begin{tabular}[c]{@{}c@{}}{0.968}  \\ {[0.962, 0.974]}\end{tabular} \\ \hline
\multicolumn{4}{l}{
\textit{Note.} The 95\% confidence interval is in each bracket. }
\end{tabular}
\end{table}

\begin{figure}[!t]
\centering
\centerline{\includegraphics[width=\linewidth]{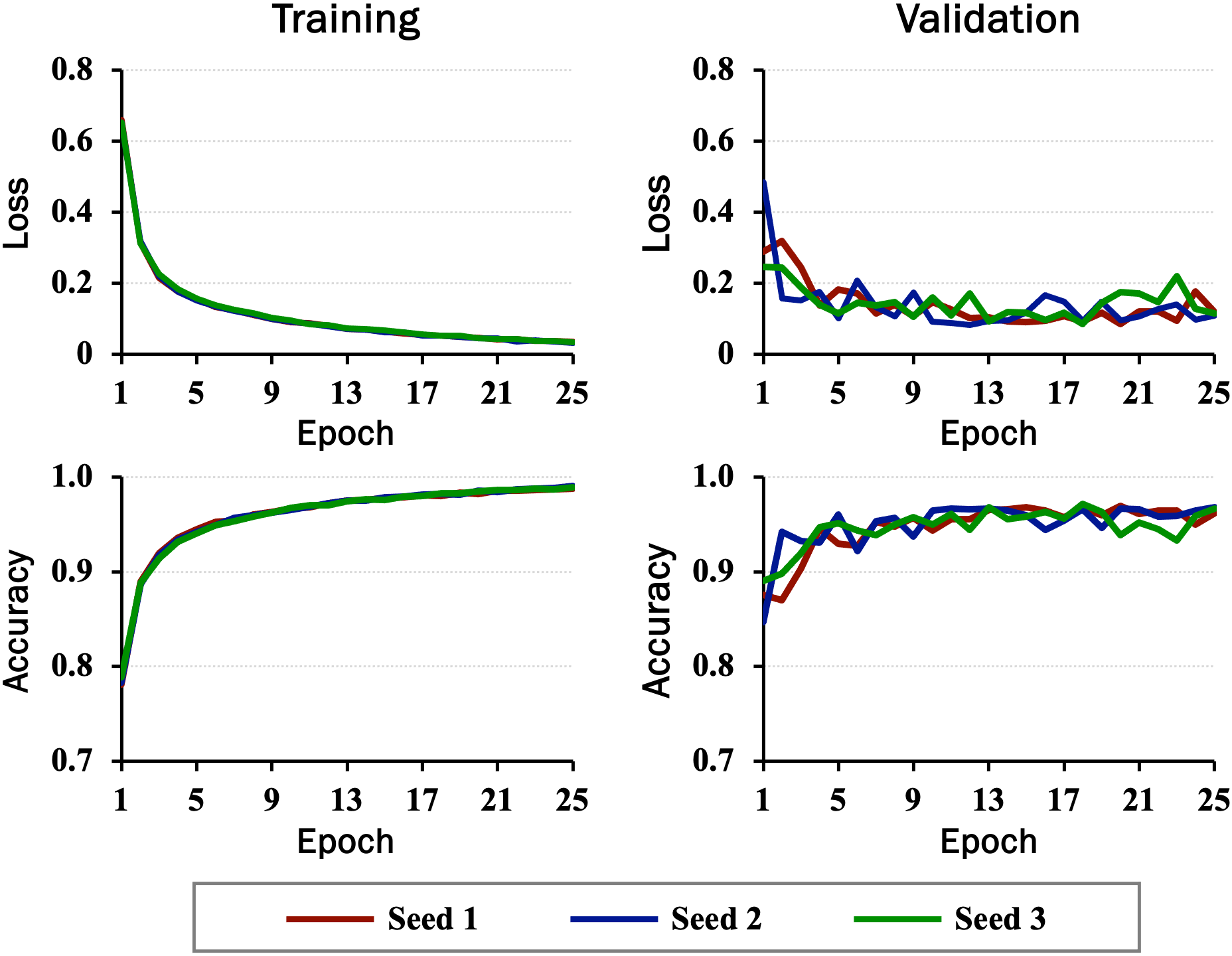}}
\caption{Performance plots of training DenseNet-121 models three times with different seeds. The first column shows the training loss and accuracy according to the training epochs, and the second column shows the validation loss and accuracy according to the training epochs. }
\label{fig: discuss_seeds}
\end{figure}

\subsection{Classification Performance Stability}
As aforementioned, our classification model with the DenseNet-121 network showed an accuracy of 97\%. To study the performance stability of the classification model, here, we compared three models training with different seeds. Specifically, we trained three DenseNet-121 networks from scratch using the NIH Siemens training dataset (n=1,363) the same as Task 1. Then, we tested the model performance on the NIH Siemens test dataset (n=313).

Table~\ref{tab:discuss_seeds} reports the classification performance for each seed model. We could observe that all three DenseNet-121 models achieved an F1-score of higher than 96\% and an accuracy of about 97\%. In addition, as shown in Fig.~\ref{fig: discuss_seeds}, when displaying plots of loss and accuracy for training and validation with respect to the training epochs, as model training continued, the training loss values of the three models decreased almost equally while the accuracy increased. These were also similarly shown in the loss and accuracy graphs of the validation, which suggests the stability of our MRI series classification network.

\add{It is remarkable that our model shows stable performance without underfitting or overfitting even though the model was trained from scratch. Compared to the prior works of 2D medical image classification \cite{kumar2016ensemble, azizi2021big} that can use pre-trained networks, as our work aims to classify 3D MRI volumes, there are no well-established pre-trained weights. In particular, for the DenseNet-121 model, the number of learnable parameters of the 3D network is about 1.6 times that of the 2D network (11M vs. 7M), which may readily lead to overfitting. However, note that we trained the model using a large amount of data: 13,606 3D MRI volumes from 1,363 studies in that each study has all eight different types of series. Also, the final prediction was estimated by ensembling the outputs of the five networks from the cross-validation. These made our model available to overcome the potential overfitting issue and achieve high stability. We expect that the overfitting problem might be further addressed by adopting various data augmentation methods for model training.  }


\begin{figure}[!t]
\centering
\centerline{\includegraphics[width=\linewidth]{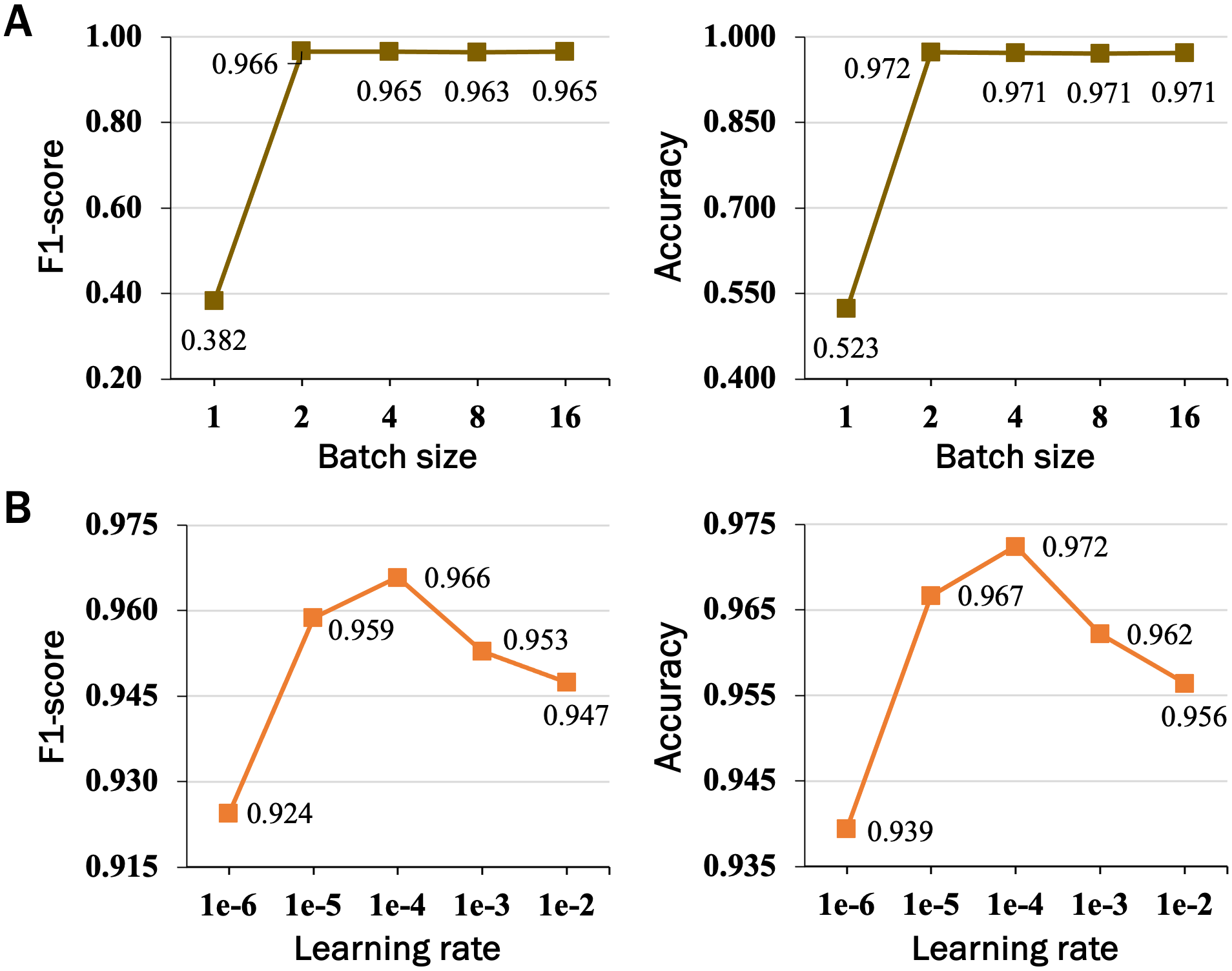}}
\caption{\add{Classification performance of the DenseNet-121 model according to the hyperparameter setting of (A) batch size and (B) learning rate. The F1-score and accuracy metrics are visualized. }}
\label{fig: discuss_hyper}
\end{figure}

\subsection{\add{Hyperparameter Setting}}
\add{In training our model, the batch size and learning rate were given as hyperparameters. To investigate the optimal value of each hyperparameter for the best classification performance, using the training and testing datasets in Task 1, we trained the DenseNet-121 model by varying each hyperparameter while keeping the others fixed, and compared the model performance.}

\add{Fig.~\ref{fig: discuss_hyper} visualizes the performance of our model according to the batch size and learning rate. Specifically, from Fig.~\ref{fig: discuss_hyper}(A) showing the performance against the batch size, we could observe that while the model showed an F1-score of 0.382 and an accuracy of 0.523 when the batch size was set to 1, the model consistently achieved over 0.96 F1-score and 0.97 accuracy when the batch size was set to larger than 1. On the other hand, in Fig.~\ref{fig: discuss_hyper}(B) displaying the performance against the learning rate, we could see that the model achieved better performance when increasing the learning rate from 1e-6 to 1e-4, but, the performance slightly decreased when the learning rate was greater than 1e-4. From these results, we set the hyperparameters of batch size to 2 and the learning rate to 0.0001 in our training strategy.}

\subsection{Limitations and Future Directions}
Although we presented a body MRI series classification model and evaluated it using various test datasets from different scanners and institutions, there are several limitations in our work. 
First, in the study of the robustness of our model to external datasets (Task 3), the model was not tested on the external MRI data acquired at the level of the chest and only tested on the different scanner data from our NIH dataset and the abdomen and pelvis data from the DLDS and CPTAC-UCEC datasets. There were no publicly available chest MRI datasets with a variety of series we have in our dataset. This may be due to the difficulty of obtaining chest MRI exams with the motions of patients.

Also, the amount of Philips scanner data available for Task 4 training was much lower than the Siemens scanner data, leading to lower performance than the result of the internal Siemens dataset. Specifically, as the number of studies from the Philips scanner was 163, the model could not be trained using more than 729 different scanner studies that were the plateau point to obtain over 96\%. This resulted in 92\% accuracy when training the model from scratch using the combination of different scanner data and 95\% accuracy when fine-tuning the model using the small number of Philips data.

In addition, our work handled the classification of only 8 axial MRI series types. While these 8 types constituted about 70\% of the series types at our institution, other series including localizers, sagittal, and coronal series were excluded. Furthermore, there may be additional series types that are acquired at other institutions that we do not account for in our study. Despite this, our work would be a pilot experiment for the commonly utilized MRI sequences across various institutions.

Moreover, the classification model did not use the DICOM header information but utilized only the 3D volume as an input. While the DICOM header that has information on MRI acquisition parameters such as the repetition time and the echo time would be potentially useful for the MRI series classification, it is often inconsistent or highly variable across scanners and institutions. Thus, we trained the model using the scanned volumes to obtain stable performance regardless of the high variance of the acquisition protocol information.

For future studies, our work can be extended to classify additional series such as coronal, sagittal, in-phase, and out-phase data. Here, data from other scanner manufacturers such as Toshiba or additional data from Philips and GE scanners would serve to improve the classification performance of the deep learning model. Also, leveraging the DICOM header elements along with voxel-level information in the model input would further enhance the performance of classifying various types of MRI series.

\section{Conclusion}
In conclusion, we presented the classification model for eight different series types of mpMRI. The model with the DenseNet-121 network performed the classification accurately on various datasets obtained from different scanners and institutions. Thus, we expect that the model can facilitate automated mpMRI hanging protocols in PACS and has the potential to reduce the reading time for radiologists and improve their workflow.

\bibliographystyle{IEEEtran}

\end{document}